\documentclass[useAMS,usenatbib]{mn2e}
\usepackage{psfig}  

\usepackage{amstext}
\usepackage{amssymb}
\usepackage{amsmath, float}
\usepackage{graphicx}
\usepackage{txfonts}

\newcommand       \be           {\begin{equation}}
\newcommand       \ee           {\end{equation}}
\newcommand       \bea          {\begin{eqnarray}}
\newcommand       \eea          {\end{eqnarray}}
\newcommand       \kms		{\,{\rm km \,\, s}^{-1}}
\newcommand       \cm		{\,{\rm cm }}
\newcommand       \msun		{\,{\rm M_\odot}}
\newcommand       \mspy 	{\,{\rm M_\odot \, yr^{-1}}}

\newcommand       \ergs		{\,{\rm erg \,\, s}^{-1}}
\newcommand       \sw		{\rm Sw 1644+57 \,}
\newcommand       \swp		{\rm Sw 1644+57.}

\newcommand       {\tff}                   {\ensuremath{t_{\rm ff}}}

\newcommand\plotone[1]
 {\centering \leavevmode \includegraphics[width={0.99\columnwidth}]{#1}}

\begin{document}

\title[The Longest Gamma-ray Burst?]{{\it Swift} 1644+57:  The Longest Gamma-ray Burst?}
\author[E. Quataert \& D. Kasen]{E. Quataert$^{1,2}$\thanks{E-mail: eliot@berkeley.edu} \& D. Kasen$^{1,2,3}$\thanks{E-mail: kasen@berkeley.edu}  \\
  $^{1}$Astronomy Department and Theoretical Astrophysics
  Center, University of California, Berkeley, 601 Campbell Hall,
  Berkeley CA, 94720\\ $^{2}$Department of Physics, University of
  California, Berkeley, Le Conte Hall, Berkeley, CA 94720 \\
  $^{3}$ Nuclear Science Division, Lawrence Berkeley National Laboratory, 1 Cyclotron Road, Berkeley, CA \\}

\maketitle

\begin{abstract}
 {\it Swift} recently discovered an unusual gamma-ray and x-ray transient (Sw
  1644+57) that was initially identified as a long-duration gamma-ray
  burst (GRB).  However, the $\sim 10$ keV x-ray emission has
  persisted for over a $\sim$ month with a luminosity comparable to
  its peak value.  The astrometric coincidence of the source with the
  center of its host galaxy, together with other considerations,
  motivated the interpretation that Sw 1644+57 was produced by an
  outburst from a $\sim 10^{6-7} M_\odot$ black hole at the center of
  the galaxy.  Here we consider the alternate possibility that Sw
  1644+57 is indeed a long-duration GRB, albeit a particularly long
  one!  We discuss the general properties of very long-duration, low-power GRB-like transients associated with the core-collapse of a
  massive star.  Both neutron star (magnetar) spindown and black hole
  accretion can power such events.  The requirements for producing low-power,  very long-duration GRBs by magnetar spindown are similar to
  those for powering extremely {\it luminous} supernovae by magnetar spindown,
  suggesting a possible connection between these two unusual types of transients.  Alternatively, \sw could be associated with the {\it faintest} core-collapse explosions:  the collapse of a rotating red supergiant in a nominally failed supernova can power accretion onto a solar-mass black hole for up to $\sim 100$ days; the jet produced by black hole accretion inevitably unbinds the outer envelope of the progenitor, leading to a weak $\sim 10^{49}$ erg explosion.    In both neutron star and black hole models, a jet can burrow through  the host star in a few days, with a kinetic luminosity  $\sim 10^{45-46} \ergs$, sufficient to power the observed emission of Sw 1644+57.
\end{abstract}

\vspace{-0.7cm}
\section{Introduction}
\label{sec:int}
\voffset=-2cm
\vspace{-0.1cm}

\sw was detected by the Burst Alert Telescope (BAT) onboard {\it
  Swift} on March 28, 2011 \citep{burrows11,levan11}.  Followup
observations with the X-ray Telescope detected a bright point
source a few hours later.  Unlike essentially all other gamma-ray
bursts (GRBs), however, \sw re-triggered BAT three additional times in the
first two days. Moreover, the X-ray emission associated with \sw has
persisted for more than a month at $L_X \sim 10^{47} \ergs$ (isotropic);
and although the emission is highly variable on timescales of minutes
to days, it is not clear that it is fading significantly
in time.  This is very different from both short and
long-duration GRBs, making \sw unique amongst extragalactic gamma-ray transients.

The host galaxy of \sw is a low mass star-forming ($\sim 0.5 \mspy$)
galaxy at a redshift of $z = 0.35$ \citep{levan11}.  There is no evidence for an
optical counterpart to the high-energy transient but the near-infrared
(NIR) flux faded by a factor of 3 over $\sim 5$ days, indicating that
the transient contributed significantly to the NIR emission, at least
at early times (when the X-ray flux was also higher).  In addition to the NIR emission, follow-up observations detected \sw in the radio, with the  flux brightening by a factor of a few in the first week.  NIR astrometry with HST and VLBA observations both determined that \sw is at the center of its host galaxy to within $\sim 0.03"$ or $\sim 150$ pc (1 $\sigma$).

The position of \sw relative to the center of its host galaxy, its
uniqueness relative to known GRBs, and the qualitative similarity
between its spectral energy distribution and those of blazars
motivated the interpretation that \sw is powered by a relativistic jet
created by accretion onto a $\sim 10^{6-7} \msun$ black hole at the
center of its host galaxy.  Moreover, the energetics of the transient,
and the reasonably strong limits on pre-outburst emission (e.g., from
ROSAT), are broadly consistent with the accretion being powered by the
tidal disruption of a solar-type star \citep{bloom11,burrows11}.

Although the tidal disruption interpretation of \sw is quite plausible, it is worth exploring alternate explanations of these unique observations.  In this {\it Letter}, we examine the possibility that \sw is in fact a new form of a long-duration GRB; by this we mean that the emission is powered by a relativistic outflow created during the core-collapse of a massive star. 
Our goal in this {\it Letter} is not to understand all of the observed properties of Sw 1644+57, but rather to assess the zeroth order plausibility of whether it could be associated with the core-collapse of a massive star.   In \S \ref{sec:grb} we assess (1) the conditions
under which neutron star spindown and/or black hole accretion can
power a very long timescale high energy transient (\S
\ref{sec:energy}) and (2) whether low-power jets from a central engine
can escape their host star or supernovae ejecta (\S \ref{sec:jet}).
We  apply these models to \sw in \S \ref{sec:sw}.
We conclude by highlighting the many outstanding questions (\S \ref{sec:discussion}).

\vspace{-0.7cm}
\section{Low-Power Gamma-ray Bursts}
\label{sec:grb}
\vspace{-0.1cm}

\subsection{Energetics and Timescales}
\label{sec:energy}

Low-power outflows (by GRB standards) during the core-collapse of massive stars can be produced by the spindown of a rapidly rotating neutron star \citep{metzger07} or accretion onto a central black hole.   A low power does not imply that the event is sub-energetic relative to canonical GRBs, only that the timescale to extract the energy is much longer.   Neutron star-powered activity would be associated with a successful core-collapse explosion while black hole accretion could be powered by the infall of the stellar envelope in a failed explosion, or the fallback of material that remains bound during an otherwise successful explosion
\citep{woosley93,macfadyen99}.

A neutron star with a spin period of $1 \, P_{\rm ms}$ ms and a
magnetic field strength of $10^{14} B_{14}$ G has a rotational energy
of $E_{{\rm rot}} \simeq 2\times 10^{52}P_{{\rm ms}}^{-2}\,{\rm
  ergs}$, a relativistic dipole spindown power of $ \dot{E} \simeq
10^{47}\, B_{14}^{2} \, P_{{\rm ms}}^{-4} \, \ergs$ and a spindown timescale of
$t_{\rm spindown} \simeq 2 \, B_{14}^{-2} \, P_{{\rm ms}}^{2} \, {\rm days}$.  Powering a month-long event with a total energy of $\sim 10^{51-52}$ ergs thus requires $P \sim 1-3$ ms and $B \sim 3 \times 10^{13} - 10^{14}$ G.  For vacuum dipole spindown, $\dot E$ is
relatively constant for $t \lesssim t_{\rm spindown}$ while for $t
\gtrsim t_{\rm spindown}$, $\dot E \propto t^{-2}$.  Note, however,
that this specific prediction for the temporal power-law index for
late-time spindown only applies for a braking index of 3, which is not
typically observed for pulsars (e.g., \citealt{kaspi06}).

\begin{figure}
\plotone{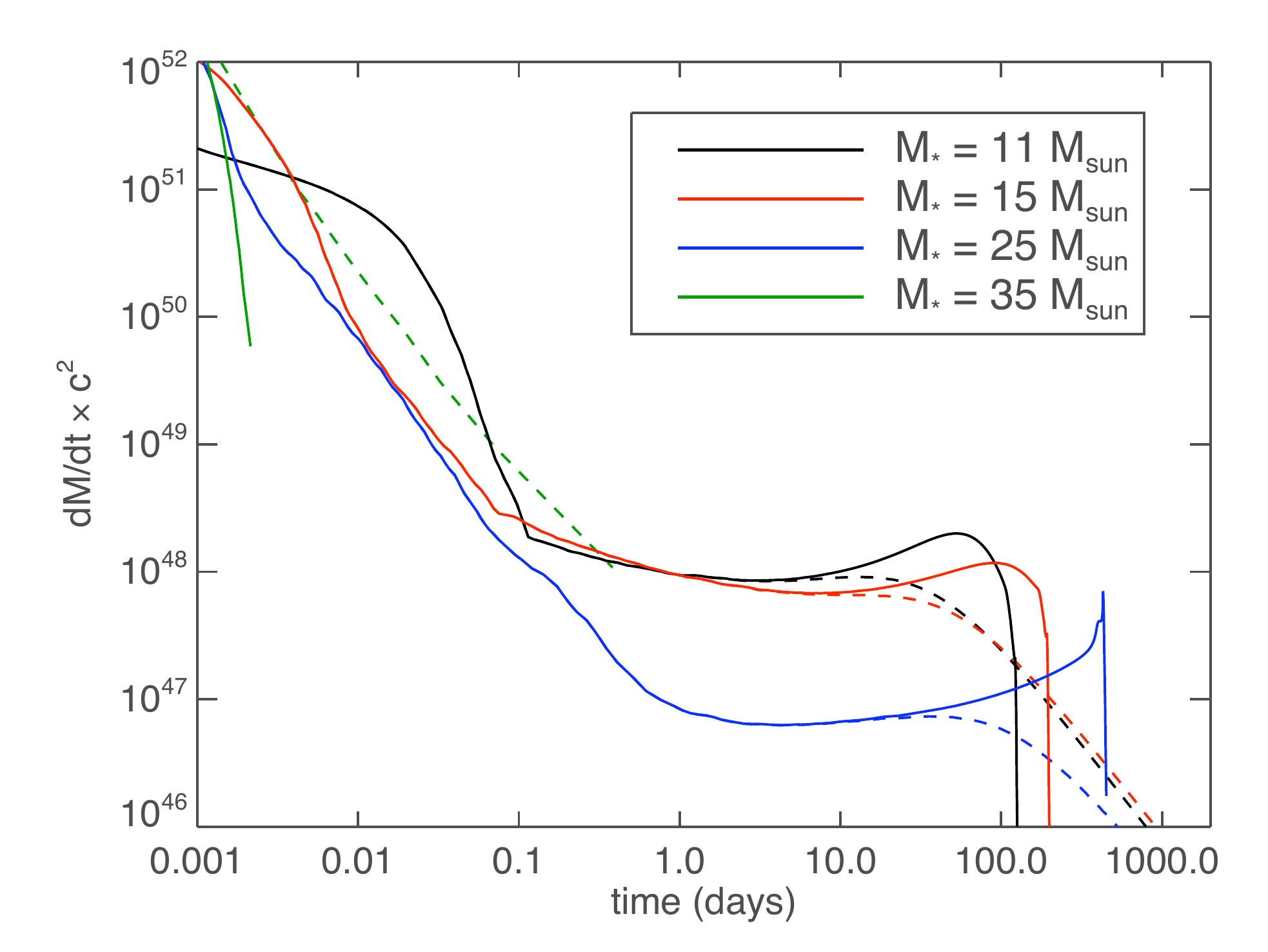}
\caption{Power available from accretion onto a central black hole for
  several pre-collapse stellar progenitors from \citet{woosley02}.  The  progenitors  with initial
  masses $M \lesssim 30~\msun$
  are red supergiants and the
  infall of the stellar envelope can power accretion for $\sim
  100-300$ days.  The solid lines assume free-collapse of the
  entire star, which is quantitatively applicable only if there
  is either no stellar explosion or a bipolar explosion in which the
  equatorial region continues to accrete.  The dashed lines show the effect of
  a supernova explosion which resulted in a linear
  velocity law, $v(r) = v_\star (r/R)$, with $v_\star$ equal to twice the
  escape velocity at the surface.  For the red supergiant progenitors, this corresponds to a weak explosion with an energy of only $\sim 10^{48}$ ergs.}
\label{fig:edot}
\end{figure}

The timescale for black hole accretion to power central engine activity
depends on the rotation and density profiles of the progenitor star
and the energy of the explosion \citep{kumar08} -- the latter because it
determines how much material remains bound to the black hole.  In the
simplest case of a failed explosion, the timescale on which
infall occurs is set by the free-fall time  
\begin{equation}
\tff(r) = \frac{ \pi r^{3/2} }{(2 G M)^{1/2} }   
\simeq 702~
~\biggl[ \frac{r}{10^{14}~{\rm cm}} \biggr]^{3/2}
\biggl[ \frac{M(r)}{10 \msun} \biggr]^{-1/2}~{\rm days}.
\end{equation}
In order to power a long timescale transient like Sw 1644+57,  a
weakly bound red supergiant (RSG) progenitor with radius $R > 10^{13}$~cm is required. 
For  a  power law density profile, $\rho(r) = \rho_0 (r/ R)^{-n}$, the enclosed mass $M(r) \propto r^{3-n}$ (for $n < 3$) and the free fall accretion rate is
\begin{equation}
\dot{M}(t) = 4 \pi \rho r^2 \frac{dr}{d\tff} = 
\frac{2(3-n)}{n}  \frac{M}{t_{\rm ff,R}}
\biggl( \frac{t}{t_{\rm ff, R}} \biggr)^{[6/n] - 1 - n},
\end{equation}
where $t_{\rm ff, R}$ is the free-fall time evaluated at the outer radius. The total power available from stellar infall is thus
\begin{equation}
\dot{E}(t) = \dot{M}c^2 \simeq  6 \times 10^{47}~M_{10}^{1/2} R_{14}^{-3/2} \biggl( \frac{t}{t_{\rm ff, R}} \biggr)^{[6/n] - 1 - n}~{\rm ergs~s^{-1}}.
\label{eq:lum}
\end{equation}
where the stellar envelope mass is scaled to $10 \, M_\odot$ and the radius to $10^{14} \, \cm \simeq 10^3 \, R_\odot$.     Presumably this energy will be  tapped with only fractional efficiency to power a jet, but depending upon the collimation the resulting isotropic equivalent power could be of order equation~\ref{eq:lum} or greater.   The actual accretion rate may deviate from the pure free-fall estimate used here since radial pressure support at small radii can slow the infall \citep{lindner10}.    In addition, the  accretion energy depends not only on the infall rate, but also on the angular momentum profile of the progenitor, since only the material that circularizes in a disk will be available to power a jet.

As a concrete example, Figure~\ref{fig:edot} shows the free fall accretion rate for the non-rotating, solar metallicity pre-supernova progenitor models of \cite{woosley02}.
The models with initial masses $\lesssim 30~\msun$  are RSGs with radii $R = 0.5 -1 \times 10^{14}$~cm, while the models with higher masses have lost their hydrogen envelope and have much smaller stellar radii ($R \sim 10^{11}$~cm).  The outer density profile for the stars with convective hydrogen layers is shallow and roughly follows a power law with  $n=2$, which gives (eq.~\ref{eq:lum})  a nearly constant accretion rate over the timescale of 200$-$300 days. The power shuts off very rapidly after this point because the stellar photosphere has been accreted.

If the star undergoes a successful supernova explosion, the accretion onto a central black hole at late times depends on how much material remains bound.   The escape velocity for a RSG is only $v_{\rm esc}(R) = 50-100~\kms$;  thus even a weak (spherical) explosion can unbind the outer hydrogen envelope and limit the late time accretion.  For the layers that do remain bound, material reaches a turnaround radius  $r_t = r/ [ 1 - v^2/v_{\rm esc}^2]$, and then falls back on a timescale $\tff(r_t)$.   In Figure~\ref{fig:edot}, we show how including a linear velocity profile of the form $v(r) = v_\star (r/R)$ modifies the late-time accretion power.\footnote{In reality, the supernova shock will also alter the density distribution of the star; this effect has been neglected here.}  For  expansion at $v_\star = 2  v_{\rm esc}(R)$,  the accretion rate remains constant until $t ~\approx 50$~days, and then declines as a power law.  For expansion velocities much larger than this,  the power drops off at yet earlier times $ <  1$~day.  Thus to explain the long duration of Sw 1644+57 within the context of black hole accretion, a RSG must have undergone the weakest of explosions, or none at all.

\vspace{-0.3cm}
\subsection{Interaction with the Host Star}

\label{sec:jet}

Standard long-duration GRBs have durations $\sim 1$ minute and are
associated with Type Ibc supernovae \citep{wb06}.  These
two facts are not unrelated: the compact stripped progenitors
associated with Type Ibc supernovae are the only progenitors in which a
jet can escape its host star on a timescale comparable to the duration
of the GRB itself \citep{matzner03}. We now consider the same
reasoning but applied to much lower power jets.

A jet with a momentum flux $\dot P_j$ has an associated kinetic power
of $L_j = \dot P_j \, v_j$, where $v_j$ is the velocity of the jet
material.  If the jet interacts with an ambient medium of density $\rho_a$,
the speed of the head of the jet through the ambient medium $v_h$ can
be estimated by considering the balance between the momentum flux of
the jet and the ram pressure of the ambient medium $\rho_a A_j v_h^2$
\citep{bc89, matzner03}, where $A_j \sim \pi \, \theta_j^2 r^2$ is the
surface area of the head of the jet and $\theta_j$ is the jet's
opening angle.  This yields: \be v_h \simeq
\left(\frac{L_j/A_j}{\rho_a v_j^3}\right)^{1/2} v_j \, \simeq
\left(\frac{L_{i}/c^3}{4 \pi r^2 \rho_a }\right)^{1/2} c 
\label{eq:vh}
\ee where $L_{i} \simeq 4 L_j/\theta_j^2$ is the isotropic equivalent
kinetic power in the jet and in the second equality we have assumed
that the jet is relativistic. Equation~\ref{eq:vh} implies $v_h \sim 0.007 \,
L_{i,48}^{1/2} \, M_{10}^{-1/2} \, R_{14}^{1/2} \, c$ where $L_{i}$ is scaled to $10^{48} \ergs$, which is appropriate for $L_j \sim 10^{45-46} \ergs$ and $\theta_j \sim 3-10$ deg.

As the head of the jet propagates through the star (and/or supernova ejecta), a cocoon of shocked stellar and jet material forms around the jet.   This cocoon in turn drives a lateral shock into the ambient medium.   The speed of this lateral shock $v_l$ can be estimated by balancing the pressure in the cocoon with the ram pressure of the lateral shock.   Since the jet produced by a central magnetar and/or black hole accretion disk is very likely to be magnetized, the same applies to the cocoon.   The toroidal magnetic field in the cocoon in turn creates an asymmetric pressure distribution, with the pressure being much larger near the jet axis than at large cylindrical radii \citep{bl92}.  This reduces the lateral expansion speed of the cocoon.   To account for this, we estimate the pressure in the cocoon that drives the lateral expansion as $p_c \simeq f E/(3 V)$ where $E$ is the total energy supplied by the central engine up to the time of interest, $V$ is the volume of the cocoon, and the factor $f < 1$ accounts for the pinching effect of the toroidal magnetic field.   The speed of the lateral shock driven by the cocoon is thus
$v_l \simeq v_h \, f^{1/4}  \, \theta_j^{1/2} \, (c/v_h)^{1/4}$.

We now consider the limit in which the timescale for the jet to escape the surrounding star is short compared to the expansion time of the stellar envelope.  This is appropriate, e.g., for a failed supernova explosion, as in the supergiant collapse scenario considered in \S \ref{sec:energy}.   In this case, the time for the jet to escape the progenitor is \be t_{esc} \simeq 5 \, L_{i,48}^{-1/2} \, M_{10}^{1/2}
\, R_{14}^{1/2} \, {\rm days} \ \ (\rm \it no \, expansion)
\label{eq:tesc1}
\ee
The corresponding lateral speed of the cocoon-driven shock is 
\be
v_l \simeq 0.3 \, v_h \, \left(\frac{f}{0.03}\right)^{1/4} \, \left(\frac{\theta_j}{3 \, {\rm deg}}\right)^{1/2} \, \left(\frac{v_h}{0.01 \, {\rm c}}\right)^{-1/4}
\label{eq:vlatnum}
\ee
where we have scaled the reduction factor $f$ to a value appropriate if the magnetic energy in the cocoon is comparable to the thermal energy (e.g., Fig.~3 of \citealt{bucciantini07}).  Equation~\ref{eq:vlatnum} implies that the lateral expansion time ($\sim [\pi/2][R/v_l]$) is a factor of $\sim 5$  longer than the time it takes the jet to escape the star, even for the low power jets of interest here.  It is thus plausible that the jet can escape the star before the cocoon completely envelops the stellar envelope.  Once the jet escapes the star, the material in the cocoon, which has a sound speed $\sim c/\sqrt{3} \gg v_l$, will escape along with the jet, depressurizing the cocoon.   After the cocoon depressurizes, the lateral shock will decelerate as it sweeps up mass, reaching a velocity of order $v_{\rm l,f} \sim (2 E_{\rm c}/M)^{1/2}$, where $E_{\rm c} \sim f L_j t_{\rm esc}$ is the  energy acquired prior to breakout.  The timescale for the lateral shock to propagate completely around the star is then\begin{equation}
t_{\rm env} \simeq 70~L_{i,48}^{-1/4} M_{10}^{1/4} R_{14}^{3/4} 
\left( \frac{f}{0.03} \right)^{-1/2}
\left(\frac{\theta_j}{3~{\rm deg}} \right)^{-1}~{\rm days}.
\label{eq:tenv}
\end{equation}
The energy of the lateral shock $\sim f L_j t_{\rm esc}$ exceeds the binding energy of the envelope of a supergiant progenitor ($\sim 10^{48}$ ergs) if $t_{\rm esc} \gtrsim 0.4 \, (f/0.03)^{-1} L_{j, 45}^{-1}$ days, where the total jet power is scaled to $10^{45} \ergs$.   This inequality also applies at each  radial shell within a given progenitor.   Thus, once the head of the jet reaches the radius $r$ where $t_{\rm esc}(r) \gtrsim 0.4 \, (f/0.03)^{-1} L_{j, 45}^{-1}$ days, the remaining outer envelope of the star is unbound, with an energy $\sim 10^{49} \, L_{i, 48}^{1/2} \, M_{10}^{1/2} \, R_{14}^{1/2} \, (\theta_j/3 \, {\rm deg})^2 \, (f/0.03)$ ergs.   For our fiducial parameters, matter is unbound outside $\sim 10^{13}$ cm.   Matter at smaller radii can, however, continue to infall onto the central black hole.  The maximum timescale over which infall can proceed without being strongly affected by the expulsion of the envelope is
\be
t_{\rm ff, max} \sim 70 \, \left(\frac{f}{0.03}\right)^{-1} \, \left(\frac{\theta_j}{3 \, {\rm deg}}\right)^{-2} \, L_{i, 48}^{-1/2} \, {\rm days}
\label{eq:maxtff}
\ee
where we have used the fact that the density profile at large radii in supergiants is  $\rho(r) \propto r^{-2}$.  These order of magnitude arguments suggest that the collapse of a RSG could potentially power jets for up to $\sim 100$ days.  One  uncertainty in these estimates is how much of the star at small radii $\lesssim 10^{10-11}$ cm falls directly into the black hole vs. circularizes in a disk; this matter can in principle produce large jet powers at early times $\lesssim 1000$ sec (Fig.~\ref{fig:edot}), which might more readily unbind the outer stellar envelope.   We have assumed that most of this mass instead forms the initial black hole.

We now consider the case of a successful stellar explosion, in which the stellar envelope expands
outwards with a velocity $v_{ej} \sim 10,000 \kms$.  In this case the
head of the low power jet initially cannot keep up with the expansion induced by
the stellar explosion.   As the stellar density decreases due to
expansion the velocity of the head of the jet increases, reaching $v_h
\sim v_{ej}$ when $R \sim (4 M c^3)/(L_i) (v_{ej}/c)^2$; using $R
\simeq v_{ej} t$ for the expanding ejecta, this implies that the jet
can escape the ejecta at a time $t_{esc}$ given by \be t_{esc} \simeq
\frac{4 v_{ej} M c}{L_i} \simeq 30 \frac{v_{ej,9} M_{10}}{L_{i,48}}
\, {\rm days} \  \ (\rm \it envelope \, expansion).
\label{eq:tesc2}
\ee where $v_{ej,9}$ is the velocity of the supernova ejecta in units
of 10,000 $\kms$.   Equation \ref{eq:tesc2} does not apply to
standard long-duration GRBs, for which the jet escape time is shorter
than the expansion time of the stellar envelope.  In the latter case
equation \ref{eq:tesc1} is the correct estimate of the jet escape time
even if the explosion is successful.

If the central engine remains active for a duration $\gtrsim t_{esc}$
then the jet can escape the surrounding stellar ejecta, potentially
powering a high energy transient.  For a black hole central engine in a supergiant progenitor, the infall time of the stellar envelope (see Fig. \ref{fig:edot}) is longer than the escape time in either a failed or weak explosion (eqs. \ref{eq:tesc1} \& \ref{eq:tesc2}).   However, because the outer envelope of a red supergiant has a binding energy of only $\sim 10^{48}$ ergs, it is easily disrupted.  In particular, equation~\ref{eq:tenv} shows that the lateral expansion of the cocoon-driven shock will eventually envelop the stellar envelope and unbind it, stifling accretion at smaller radii.  For sufficiently collimated and/or magnetized jets, however, accretion fed by the infall of the stellar envelope can last for a duration approaching the free-fall time of the outer envelope $\sim 100$ days (eq.~\ref{eq:maxtff}).

For a spinning down magnetar, the requirement that $t_{esc} \lesssim
t_{spindown}$ in a successful explosion can be shown to imply that the magnetar jet must be
sufficiently collimated in order to escape the surrounding ejecta
while most of the spindown power remains: $\theta_j \lesssim
10 \, (v_{ej,9} M_{10})^{-1/2} \, {\rm deg}$.   Initially, most of the energy flux in a neutron star outflow is in
the equatorial plane of the rotator.  However, the outflow's toroidal
magnetic field builds up outside the termination shock and can
collimate the outflow into a jet along the polar axis
\citep{bucciantini09}.  Numerical simulations in the long-duration GRB
context yield collimation angles $\lesssim 10$ deg.   However, the degree of
collimation, i.e., $\theta_j$, depends sensitively on the magnetization in the region between the termination shock and the bulk of the supernova ejecta (as in the cocoon dynamics described above eq.~\ref{eq:tesc1}).  This is difficult to predict with certainty.   If the
jet cannot escape the stellar ejecta it is plausible that the spindown
power of the magnetar is instead thermalized, heating the ejecta and
potentially powering an ultra-luminous supernova \citep{kasen10}.
Even if the jet does escape, the fact that the escape and spindown times
are comparable suggests that some of the spindown power is likely to
be transferred to the stellar envelope, contributing to the luminosity
of the supernova.

\vspace{-.8cm}
\section{Application to \sw}
\label{sec:sw}
\vspace{-0.15cm}

The zeroth-order observational requirements for explaining \sw are that a source must produce a relativistic outflow with appreciable power for several weeks with a total energy budget $\sim 10^{50-52}$ ergs; the energy  is only loosely constrained because of
uncertainties in the beaming.    \citet{burrows11} argued that the x-ray lightcurve in the first $\sim$ 3 weeks could be broadly reproduced by the $t^{-5/3}$ scaling expected for fallback and/or tidal disruption, but this fading is uncertain and depends  on how the luminosity in the observed bands is related to the bolometric luminosity.

The results of \S \ref{sec:grb} demonstrate that both a spinning down neutron star with P $\sim$ a few ms and $B \sim 3 \times 10^{13}$ G and accretion onto a newly formed solar mass black hole can match the energetics and timescale of \swp   \ The reason that \sw is so distinct from more typical long-duration GRBs is, however, fundamentally different in the neutron star and black hole models.   In the neutron star model, the required magnetic field strength for \sw is $B \sim 3 \times 10^{13}$ G rather than $B \sim 10^{15-16}$ G as in long-duration GRB models.  This increases the spindown timescale by $\sim 4-5$ orders of magnitude.   By contrast, in the black hole accretion context, the key difference between \sw and standard GRBs would be the stellar progenitor:  a red supergiant for \sw versus stripped envelope progenitors for typical long-duration GRBs.

The time it takes the low-power jet associated with \sw  to escape the stellar envelope is significantly longer than in typical  long-duration GRBs.   Nonetheless, given plausible jet powers $L_j \sim 10^{45-46} \ergs$ and the uncertainty in the collimation, the escape timescale could be as short as a few days (eqs \ref{eq:tesc1} \& \ref{eq:tesc2}).    This is true even for a supergiant progenitor which has a radius of $\sim 10^{14}$ cm.  In the context of a successful stellar explosion, the radius of the supernova ejecta at the time of jet 'breakout' would also be $\sim 10^{14}$ cm (even if the progenitor is initially much more compact).   It is possible that the $\sim$ few day timescale for the jet to escape imprints itself on the observed lightcurve, accounting for the initial few day peak of activity observed from \swp \,   

The constraints on the Lorentz factor of \sw are not very stringent (perhaps $\Gamma \sim 3-10$) relative to typical GRBs because  the emission  is  comparatively soft  \citep{bloom11}.   In the core-collapse context it is possible that \sw would be less relativistic than normal GRBs because of additional mixing with the stellar material as the low power jet traverses the star. 

In addition to the energetics and duration constraints, \sw showed significant variability  on timescales of $\sim$ 100~sec, which \citet{bloom11} associated with the dynamical time
around the event horizon of a $\sim 10^6 \, M_\odot$ black hole.   This interpretation is very plausible, but it may not be unique.  For a solar-mass central engine, one would also  expect variability on much shorter timescales, down to milliseconds.  The signal to noise in the \sw x-ray light curve is not, however, sufficient to constrain significant variability on $\lesssim 10$~sec (N. Butler, private communication).   Moreover, the temporal power spectrum of the {\it Swift} lightcurve does not show any feature at a particular timescale (e.g., $\sim 100$ sec) and is instead consistent with a power-law that reaches the noise floor for $\lesssim 10$ sec (see Fig. S1b of \citealt{bloom11}).   Given that GRBs, AGN, and X-ray binaries all have roughly power-law temporal power spectra, it is not clear that the variability of \sw clearly favors one central engine over another.  More quantitatively comparing the temporal power spectrum of \sw with these other classes of objects would be very interesting.   On the theoretical side, the longer  timescale ($\gtrsim 0.1-1$ sec) variability in canonical long-duration GRBs may arise primarily due to interaction with the surrounding star (e.g., \citealt{bucciantini09,morsony11}).   We would expect the same to be true in the context of Sw 1644+57, although the precise timescales produced by this interaction are likely to change because of the lower jet power and the different progenitor.

\vspace{-0.7cm}
\section{Discussion}
\label{sec:discussion}
\vspace{-0.15cm}

We have argued that models with central engines like those of long-duration GRBs -- solar-mass compact objects formed during the core-collapse of a massive star -- can explain the broad properties of the unusual gamma-ray transient \swp   \, Specifically, models with solar-mass compact objects 
produce jets with similar kinetic power and timescales to those invoked in the context of massive black hole accretion in \citet{bloom11} and \citet{burrows11}.    The phenomenology of the resulting emission depends largely on the properties of the jet and thus should in many ways be independent of the central nature of the engine, complicating the interpretation of \swp

The localization of \sw to near the center of its host galaxy is highly suggestive of AGN activity, but it is also not unreasonable to suspect that a stellar explosion might occur in the galactic nucleus, perhaps associated with circumnuclear star formation.   Long-duration GRBs have a tendency to appear in the brightest star forming regions of a galaxy \citep{fruchter06},  which in this case coincides with the center.  The offset distributions for GRBs constructed by \cite{bloom02} indicate a $\sim 10\%$  probability of finding a GRB within the radius allowed by observations of \sw (i.e., within $\sim 20\%$ of the galaxy half light radius).   GRB 021004, for example, was located similarly close ($< 119$~pc) to its host galaxy center \citep{fynbo05}.  

In some ways,  \sw  did not show the expected signatures of a tidal disruption event.  In the usual picture, the fallback of bound material forms a disk near the tidal disruption radius and radiates primarily in the ultraviolet/optical \citep{ulmer99}.   For systems with super-Eddington fallback rates, which are probably the most likely to power relativistic jets, a particularly bright optical transient is expected associated with outflows driven by radiation pressure \citep{strubbe09}; there are indeed several recent tidal disruption candidates selected on such optical emission \citep{farrar11, cenko11}.   The fact that no such optical transient was seen for \sw could be the result of significant dust extinction ($A_v \sim 10$~mag) in the host galaxy nucleus, which \citet{levan11} argue is consistent with the high hydrogen column density determined from the x-ray spectrum.  In the magnetar GRB model, high dust extinction would likely also need to be invoked to explain the non-detection of a supernova.  On the other hand, in the RSG collapse model the absence of a bright optical transient is to be expected given the failure (or extreme weakness) of the supernova explosion.

Continued monitoring of \sw should  help clarify its origin.  A prolonged phase of relatively constant x-ray luminosity becomes, at some point, difficult to reconcile with a tidal disruption model. Rather, one expects to see the power decline on the fallback timescale, which is $t_{fb} \sim 20 \, (M_{BH}/10^6 M_\odot)^{5/2} \, (R_p/ 3 R_s)^3 \, {\rm min}$ for a solar type star  (where $M_{BH}$ is the black hole mass and $R_p$ is the peribothron distance of the stellar orbit, scaled to 3 Schwarzschild radii).   For a $10^6 \, (10^7) \, M_\odot$ black hole, $t_{fb} \lesssim 7 \, (20)$ days unless the disrupted star is a giant with a large radius.    For $t \gtrsim t_{fb}$, the jet power should decrease in time, which is not readily apparent in the recent {\it Swift} data for \sw (though the interpretation is complicated by the difficulty of relating the luminosity in the {\it Swift} bandpass to the bolometric luminosity, let alone to the jet power or accretion rate).   In the magnetar model,  the jet power will remain roughly constant for the initial spindown timescale of the neutron star. Depending on the jet collimation and efficiency, the requisite power can be maintained for significantly longer than a month while still satisfying the energy constraints of a maximally spinning neutron star (\S \ref{sec:energy}).  The RSG collapse model predicts a nearly constant jet power for up to $\sim 100$ days, followed by a rapid drop off (see \S \ref{sec:jet} and Fig. \ref{fig:edot}).

If \sw was in fact of stellar origin, one might ask why its properties were so  discontinuous compared to any other GRB observed to date.  The magnetar model provides no obvious explanation  -- presumably a continuous range of magnetic field strengths, and hence spin down rates, could be realized.  In the supergiant collapse case, on the the other hand, the discontinuity reflects the bimodality of progenitor radii  depending on whether or not a massive star retains its  hydrogen envelope.   Figure~\ref{fig:edot} demonstrates that this bimodality is in fact predicted in the set of \cite{woosley02} progenitors of varying masses.   The low rate inferred from \sw  suggests that, compared to stripped envelope stars,  collapse and relativistic jet production in RSG progenitors is a rare event, if it happens at all.  

The large energy injection on week-month timescales required to understand \sw is similar to the energy injection required to power ultraluminous supernovae \citep[e.g.,][]{quimby07,miller09} by magnetar spindown \citep{kasen10}.   Moreover, within the (very large) uncertainties, the rate of ultraluminous supernovae \citep{quimby09} is comparable to the estimated rate of events like \swp \, It is thus possible that these seemingly different transients are closely related, with neutron star spindown being the central engine in both cases.  

If, on the other hand, \sw was powered by black hole accretion from stellar collapse, the outburst may  be associated with some of the {\it faintest} supernovae known.   Observations of Type~IIP supernovae indicate that the explosion energy achieved in the core collapse of RSGs varies significantly from case to case \citep{hamuy03}, including several recorded instances of  very weak mass ejections \citep[$E < 10^{50}$~ergs;][]{zampieri03, pastorello04, fraser10}.  The luminous red novae are even dimmer transients with inferred explosion energies of order the binding energy of a RSG $\sim 10^{48}$~ergs \citep{kulkarni07,thompson09,bond09}.   Pre-explosion images of some luminous red novae suggest that the progenitors are relatively massive stars ($M \sim 10~\msun$) heavily enshrouded in dust \citep{prieto08}.  It remains unclear, however, whether these events represent the true core collapse of a star or rather just a pulsational episode that unbinds some of the hydrogen envelope.  In any case, given the range of observed outcomes, it seems possible that in some rare circumstances the supernova shock in a RSG envelope might only reach a few times the escape velocity, or fail to develop altogether.  Provided the progenitor had sufficient angular momentum, a likely outcome appears to be a GRB of low power and unusually long duration, similar to \swp

\vspace{-0.8cm}
\section*{Acknowledgments}
\vspace{-0.1cm}
We are grateful to J.~Bloom, N.~Butler, B.~Cenko, B.~Metzger, D.~Perley, and E.~Ramirez-Ruiz for very helpful discussions.  This research has been supported by the DOE SciDAC Program (DE-FC02-06ER41438).  EQ was supported in part by the David and Lucile Packard Foundation.  \vspace{-0.7cm}
\bibliography{refs}


\end{document}